\documentclass[aps,prb,twocolumn,groupedaddress]{revtex4}
\usepackage{graphicx}
\begin{document}
\title{Joint density-functional theory for electronic structure of solvated systems}
\author{S.A.~Petrosyan}
\affiliation{Laboratory of Atomic and Solid State Physics, Cornell University}
\author{Jean-Francois Briere}
\affiliation{Laboratory of Atomic and Solid State Physics, Cornell University}
\author{David Roundy}
\affiliation{Laboratory of Atomic and Solid State Physics, Cornell University}
\author{T.A.~Arias}
\affiliation{Laboratory of Atomic and Solid State Physics, Cornell University}
\date{\today}

\begin{abstract}
We introduce a new form of density functional theory for the {\em ab
initio} description of electronic systems in contact with a molecular
liquid environment.  This theory rigorously joins an
electron density-functional for the electrons of a solute with a
classical density-functional theory for the liquid into a single
variational principle for the free energy of the combined system.  
A simple approximate functional predicts, without any
fitting of parameters to solvation data, solvation energies as well as
state-of-the-art quantum-chemical cavity approaches, which require such
fitting.
\end{abstract}


\maketitle

{\em Ab initio} electron density-functional methods have proved a
computationally efficient and accurate tool for the exploration of a
wide range of issues in condensed-matter physics and chemistry.  (See, for instance \cite{payne92,seminario96}.)
However, application of this
approach is largely limited to the solid-state, gas phase
chemistry, or to surface chemistry in high vacuum environments, with
practical problems involving liquid environments largely out of reach.
The reason for this unfortunate limitation is that proper
description of the effects of water demands not only the inclusion of
many solvent molecules but also thermodynamic sampling of many
configurations of those molecules so as to properly capture the
structure and response of the liquid over experimental length and time
scales.  Thus, the applicability of such
approaches to the vast array of problems involving liquid environments,
from liquid fuel-cell research to biochemistry, is severely curtailed.

In response, much effort (many thousands of articles and a large
number of reviews, \cite{tomasi94,tomasi97,rivail95,cramer99,luque03}
and others, in the last decade alone) has been dedicated to the
development and application of a large number of different
``continuum'' solvation models, in which the details of the molecular
aqueous environment are replaced by a continuum description.  More
recently, Dzubiella et al.\cite{Dzubiella} developed a continuum model for
water where hydrophobic, dispersion and electrostatic energy terms are
written as functionals of the exclusion volume.
The plethora of models evidences the importance of the problem, but
the lack of a consensus underscores that no truly satisfactory method
has been found.

The weaknesses of the current state of the art in continuum solvation
approaches arises from their basic structure.
As Ref.~\cite{luque03} and the other reviews cited above describe,
such approaches generally divide the free energy of solvation into a
number of contributions, typically cavitation (formation of the
solvent-solute interface), dispersion (long-range electron
correlations), repulsion (short-range electron overlap effects), and
electrostatic (reorientational and polarization screening in the
solvent).  Work then proceeds to develop models for each of these
terms separately.  The models for these terms generally require a
cavity shape, which is most usually defined by spheres representing
atoms or functional groups, where the sphere radii are determined by
fits of the final results to a large {\em empirical} database of
solvation energies\cite{tomasi94}.
Finally, to account for nonlinear saturation effects near solutes, an
intermediate dielectric constant is often used in a shell around the
solute as a buffer between the solute and bulk medium, as in
Ref.~\cite{sat} for example.  Despite the successes of this method,
the {\em ad hoc} separation of physical effects (all originating
ultimately from the underlying quantum and statistical mechanics) and
the empirical fitting to precisely the class of quantities of interest
limits the reliability of the predictions of these models.  This is
especially true for applications to new classes of chemical species or
to situations outside of the fitting database, such as to study liquid
phase surface catalysis.

The new approach which we propose to pursue here suffers none of the
aforementioned difficulties.  We first prove a theorem which shows
that the thermodynamics of a system and its electrons ({\em solute}) 
in equilibrium with a liquid environment ({\em solvent}) can be
described rigorously in terms of a {\em joint density-functional
theory} (JDFT) between the electrons in the system and the molecules
comprising the solvent.  The physics of the equilibrium between a
solute and a solvent (cavity formation, dielectric screening,
dispersion and repulsion) are then all determined in a single
variational principle.  Maintaining the first principles nature of
density-functional theory, this new approach thus requires no
artificial separation of contributions, no {\em ad hoc} definitions of
cavity shapes, and no empirical fitting of parameters to experimental
solvation energies.  As shown below, even a preliminary implementation
of this new approach gives results which are competitive with
state-of-the-art continuum solvation models, even without direct fitting to any solvation data.  This suggests that further refinements will
result in a new,
efficient and predictive approach to electronic structure
in the presence of liquid environments.

{\em Joint density-functional theory ---} 
A straightforward combination of Mermin's non-zero temperature
formulation of density-functional theory\cite{Mermin65} with Capitani
{\em et al.}'s extensions of the zero-temperature theory to include
nuclear degrees of freedom\cite{Capitani} leads to the following,
exact variational principle for the total thermodynamic free energy of
an electron-nuclear system in a fixed external electrostatic potential
$V(r)$
\begin{eqnarray} 
A&=&\min_{n_t(r), \{N_i(r)\}}\left\{ F[n_t(r),\{N_i(r)\}] + \vphantom{\left(\sum_a\right)} \label{eq:DF1} \right. \\
&&  \int d^3r\,\, V(r)  \left. \left( \sum_i
Z_i N_i(r) - n_t(r) \right)\right\}, \nonumber 
\end{eqnarray}
where $n_t(r)$ is the thermally and quantum mechanically averaged
total single-particle number density of electrons, $N_i(r)$ is the likewise
averaged density of the nuclear species $i$ (of atomic number $Z_i$),
and $F$ is a universal functional.  (Here, as throughout this work, we
employ atomic units, in which Planck's constant and the mass and
charge of the electron all have value unity, $\hbar=m_e=e=1$.)  The
universality properties of the functional $F$ may be seen directly
from its construction within Levy's constrained search
procedure\cite{Levy79},
\begin{eqnarray} 
&& F[n_t(r),\{N_i(r)\}]  \equiv  \label{eq:F} \\
&& \min_{\hat \rho \rightarrow \left[ n_t(r),
    \{N_i(r)\} \right] }  
\mbox{Tr\,\,} \left( \hat \rho \hat H 
+ k_B T \hat \rho \ln \hat \rho \right),  \nonumber
\end{eqnarray}
where $k_B T$ is the thermal energy, $\hat H$ represents
all interactions among and the kinetic energy of the 
electrons and nuclei, $\hat \rho$ is the full
quantum-mechanical density matrix for the electron and nuclear degrees
of freedom, and the minimization is carried out over only those $\hat
\rho$ which lead to the given densities $n_t(r)$ and $\{N_i(r)\}$.
From this construction it is clear that $F$, like $\hat H$ from
which it derives, is independent of the external potential $V(r)$ and
depends only upon the identities of the nuclear species $i$ (and,
implicitly, upon the temperature $T$), as Capitani et
al.\cite{Capitani} found previously for the case of $T=0$.

To employ the variational principle Eq.~(\ref{eq:DF1}) in the study of
a system to be treated explicitly in contact with a solvent
environment, we take the nuclear species $i$ to be those comprising
the environment (solvent) and the potential $V(r)$ in Eq.~(\ref{eq:DF1})
to be that arising from the nuclei of the explicit system
(solute),
\begin{equation} \label{eq:Vs}
V(r)\equiv \sum_I\ Z_I/|r-R_I|,
\end{equation}
which we take to sit at fixed locations $R_I$ with atomic
numbers $Z_I$.  Note that, although we employ a Born-Oppenheimer
approximation for the nuclei of the explicit system, at this stage the
treatment of the nuclear species of the environment in Eq.~(\ref{eq:F}) is
fully quantum mechanical.  Thus, Eqs.~(\ref{eq:DF1},\ref{eq:F})
account for all zero-point motion effects associated with lighter
nuclear species in the solvent, such as may be associated with the
protons in liquid water or with the helium atoms in superfluid helium
when used as a solvent.

Although Eqs.~(\ref{eq:DF1},\ref{eq:F}) give a rigorous continuum
treatment of the environment nuclei, the variational principle
Eq.~(\ref{eq:DF1}) is ultimately impracticable because it requires
explicit treatment of all of the electrons, including those associated
with the environment.  We thus ``integrate out'' the electrons associated
with the environment by writing $n_t(r)=n(r)+n_e(r)$, where
$n_e(r)$ is the electron density associated with the environment and
$n(r)$ is the electron density associated with the system in contact
with that environment.  We then perform the minimization over all
allowable $n_e(r)$, and finally perform the minimization over all
allowable $n(r)$.  For this purpose, we define the sets of allowable
$n_t(r)$, $n_e(r)$ and $n(r)$ to be all N-representable functions
satisfying the criteria of Gilbert\cite{Gilbert75} and integrating to
the appropriate number of electrons for the respective system.
Because all thus
defined $n_t(r)$ can be constructed as the sum of some allowable
$n_e(r)$ and some allowable $n(r)$ and because all such allowable
$n_e(r)$ and $n(r)$ sum to an allowable $n_t(r)$, this procedure is
guaranteed to recover the final free energy in Eq.~(\ref{eq:DF1}).
Thus, we have
\begin{eqnarray} 
A & = & \min_{n(r), \{N_i(r)\}}\left\{ G[n(r),\{N_i(r)\},V(r)] \vphantom{\int} \right. \label{eq:DF2} \\
&& \left. - \int
d^3r\,\,V(r) n(r)\right\} \nonumber
\end{eqnarray}
where $V(r)$ is defined above in Eq.~(\ref{eq:Vs}) and where
\begin{eqnarray} 
&& G[n(r),\{N_i(r)\},V(r)] \equiv \label{eq:G} \\
&& \mbox{\ \ } \min_{n_e(r)} \left\{ F[n(r)+n_e(r),\{N_i(r)\}]  \vphantom{\sum_I \int} \right. \nonumber \\
&&\left. +\int d^3r\,\, V(r) \left(\sum_i Z_i N_i(r) -n_e(r)\right) \right\} \nonumber
\end{eqnarray}
is universal in the sense that its functional form, like $F$ from
which it derives, depends only on the nature of the solvent and,
implicitly, the temperature, and that its dependence on the solute is
only through the electrostatic potential of the nuclei in $V(r)$.  The
choice to separate the interaction $-\int d^3r\,V(r)n(r)$ in
Eq.~(\ref{eq:DF2}) from the definition of $G$ limits the
interactions which the unknown functional $G$ must describe, easing
the task of finding good approximations.  Note, for instance, that
with this choice $V(r)$ in Eq.~(\ref{eq:G}) now couples to a neutral
charge distribution, thereby limiting to the greatest extent
possible the dependence of $G$ on $V(r)$.

Eq.~(\ref{eq:DF2}) gives the exact free energy and exact configuration of
the solvent $\{N_i(r)\}$.  However, care must be taken in the interpretation
of the $n(r)$ which yield the minimum value.  The indistinguishability of
electrons implies that there can be no fundamentally meaningful assignment
of electrons as belonging either to the environment or to the system, and
thus no exact formulation can give a unique result for $n(r)$ without some
additional prescription. Indeed, for the exact $\{N_i(r)\}$ and any $n(r)$
which integrates to the correct number of electrons and is everywhere less
than the exact solution $n_t(r)$ so that $n_e(r)=n_t(r)-n(r)$ is
allowable in the above sense, the minimization in Eq.~(\ref{eq:G}) will find
$n_e(r)=n_t(r)-n(r)$ and thus ultimately produce the exact value for $A$.
There is thus a large set of $n(r)$ which yield the minimum value in
Eq.~(\ref{eq:DF2}), and the variational principle embodied in
Eqs.~(\ref{eq:DF2},\ref{eq:G}) satisfies the fundamental condition of not
enforcing any particular, arbitrary decomposition of the total electron
density into solvent and environment contributions.

In practice, however, we expect approximations to Eq.~(\ref{eq:G}) to
break the above degeneracy and to pick out a unique solution.  The
standard pseudopotential method, which replaces the effects of the
nuclei and (relatively) inert core electrons of a solid or molecule
with an effective or ``pseudo-'' potential\cite{payne92}, is in fact an
example of an approximation which tracks only a portion of the total
electron density and provides results approaching chemical accuracy
while suffering no pathologies related to the underlying degeneracy of
an apportionment of electrons between two subsystems.

The existence and reliability of so called ``molecular
pseudopotential'' Hamiltonians\cite{Vaidehi92,Kim96,Park01} implies
the existence of reliable approximations to Eq.~(\ref{eq:G}) which
pick out a unique solution for $n(r)$.  Such pseudopotential
Hamiltonians replace the effects of the nuclei and electrons of a
collection of molecules on the electrons of an external system
(solute) with an effective potential $V_{\{R_i\}}(r)$, which depends
explicitly on the locations of the molecular nuclei $\{R_i\}$.  Such
Hamiltonians have proved to be quite accurate.  Using them, for
instance, Vaidehi {\em et al.}  find the solvation energy of $Li^{+}$
to within 0.6~kcal/mole, and Kim, Park and co-workers find results for
total energies with an accuracy acceptable for the study of the
problem of an excess electron solvated in water.

Formulating the exact thermodynamics of such Hamiltonians with the
same approach that leads to Eq.~(\ref{eq:DF1}) gives directly the
principle in Eq.~(\ref{eq:DF2}), but now with
\begin{eqnarray} 
&& G[n(r),\{N_i(r)\},V(r)] \equiv \label{eq:v2}\\
&&\min_{\hat \rho \rightarrow \left[ n(r),
    \{N_i(r)\} \right] } \mbox{Tr\,\,} \left( \hat \rho \hat H_{\{R_i\},\{R_I\}} 
+ k_B T \hat \rho \ln \hat \rho \right), \nonumber
\end{eqnarray}
where $n(r)$ represents the electron density associated with the
solute alone and $\hat H_{\{R_i\},\{R_I\}}$ represents the internal
electron gas Hamiltonian for the solvent electrons {\em alone}, the
interaction of these electrons with the molecular pseudopotential
$V_{\{R_i\}}(r)$ and a model potential function $U(\{R_i\},\{R_I\})$
describing the interactions among the environment molecules and the
interaction between those molecules and the nuclei of the solute
through the electrostatic potential $V(r)$ defined in Eq.~(\ref{eq:Vs}).
Because the electrons have been already apportioned between the solute
and the solvent during the construction of $\hat H_{\{R_i\},\{R_I\}}$,
the functional $G$ in Eq.~(\ref{eq:v2}) represents an example of an
approximation to Eq.~(\ref{eq:G}) which is both reliable and free of
any potentially pathological issues associated with degenerate
solutions for $n(r)$.

With the functional dependence of $G$ established in
Eq.~(\ref{eq:DF2}), we next separate out known components and leave
an unknown part to be approximated,
\begin{eqnarray}
&& G[n(r),\{N_i(r)\},V(r)] \equiv A_{KS}[n(r)] \nonumber \\
&&\mbox{\ \ } + A_{lq}[\{N_i(r)\}] +U[n(r),\{N_i(r)\},V(r)], \label{eq:JDFT1} 
\end{eqnarray}
where $A_{KS}[n(r)]$ is standard universal Kohn-Sham
electron-density functional of the explicit system when in isolation,
$A_{lq}[\{N_i(r)\}]$ is the ``classical'' density-functional for the
liquid solvent environment when in isolation, and
$U[n(r),\{N_i(r)\},V(r)]$, defined formally and exactly as the
difference between the exact functional and the sum of the two former
functionals, is a new functional describing the coupling between the
systems.  The new functional $U[n(r),N(r),V(r)]$ has the same
universality properties as the functional $G$ from which it
derives.

{\em Construction of approximate functionals ---} Working with
(\ref{eq:JDFT1}) requires an approximate functional
$A_{lq}[\{N_i(r)\}]$ for the bulk liquid.  For water this is an active
area of research\cite{sun01,Meister,Ramirez}.  To describe water in
our preliminary implementation, we built on the ideas and perspective
of Sun\cite{sun01}.  We first imagine minimizing over the proton
density so that a single field remains, the density $N(r)$ of the
oxygen nuclei, which one may view as the ``molecular density'' as
determined by taking the oxygen nucleus to define the location of each
molecule.  Our version of the resulting functional then takes the form
\begin{eqnarray}
&& A_{lq}[N(r)] = \label{eq:Alq}\\
&& A_{id}[N(r)]+\int
N(r)\left[\epsilon_{hs}(N(r))-aN(r)\right]\,d^3r \nonumber \\
&& -b \int N(r) \left[ \int g_\sigma(r-r')N(r')\,d^3r' -
  N(r)\right]\,d^3r. \nonumber
\end{eqnarray}
The first term in Eq.~(\ref{eq:Alq}), $A_{id}[N(r)]=k_B T \int N(r)
\left( \ln\left(N(r) \lambda^3\right)-1\right)\,d^3r$ is the
analytically {\em exact} functional for the ideal gas, where $k_B T$
is the thermal energy and $\lambda$ is the thermal de Broglie wave
length of the solute molecules.  In the second term, $\epsilon_{hs}
\equiv k_B T\left((3/2)
\left((1-\eta)^{-2}-1\right)-\ln\left(1-\eta\right)\right)$ is
the Percus-Yevick approximation\cite{percus57} for the free energy per
particle of a system of hard spheres of diameter $d$ over and above
that of the ideal gas, where $\eta \equiv (\pi d^3/6) N(r)$ with $d$ being
the sphere diameter (fit to experimental parameters below).  (In
retrospect, this work should have employed the Carnahan-Starling
approximation, which more accurately describes hard spheres and
represents little or no computational overhead.  For the range of
parameters relevant here, the two approximations agree to within about
4\%, and so we expect little significant change to the final results.)
The constant $a$ in this second term describes the
cohesive tendency between molecules that holds the fluid together.
The third term is written so that it is non-zero only if $N(r)$ is not
constant, where $b$ is a coupling constant and $g_\sigma(r)\equiv
\exp(-r^2/2\sigma^2)/(2 \pi \sigma^2)^{3/2}$ is a normalized Gaussian
describing the range of non-local behavior.

The first two terms in Eq.~(\ref{eq:Alq}) capture the properties of
the bulk fluid.  The two parameters $a$ and $d$ in these terms were
adjusted to reproduce the equilibrium density and bulk modulus of
water with results summarized in Table~\ref{tbl:fits}.  For the final
term which describes inhomogeneities, the non-local coupling strength
$b$ and the range $\sigma$ were adjusted to reproduce the macroscopic
surface tension of water $\gamma$ and the approximate location $R_b$
of the ``bend'' (as measured by the point of maximum downward
curvature) of the surface tension versus sphere radius prediction of
the molecular dynamics data of ten~Wolde and
co-workers\cite{tenWolde65}.  Table~\ref{tbl:fits} summarizes these
results as well.  We emphasize that although some of the parameter
fits in this preliminary formulation are empirical, they do not
involve any direct fit to molecular solvation free energies.

\begin{table}
\begin{tabular}{l||cc|cc}
parameters& a  & d & b  & $\sigma$ \\ 
&  ($\frac{\mbox{J m$^3$}}{\mbox{mole$^2$}}$) &  (nm) &  ($\frac{\mbox{J m$^3$}}{\mbox{mole$^2$}}$) & (nm)\\ 
&&&&\\ \hline
values &
0.3944 &
0.2918 &
0.1561 &
0.4388\\ \hline \hline
properties & $N_b$  & $B$  & $\gamma$  & $R_b$ \\
& (kg/m$^3$) & (GPa) & (mJ/m$^2$) & (nm)\\
&&&&\\ \hline
this work &998.3 & 2.184 & 71.93 & 3.70\\
experiment(*) & 998.2 & 2.190 & 72.75& $\sim$4
\end{tabular}
\caption{Fit parameters from (\ref{eq:Alq}) with comparisons between model
and experimental results (at standard conditions of 20$^\circ$C and atmospheric pressure. (*) The value for $R_b$ is theoretical\cite{tenWolde65}.  (See text.)}
\label{tbl:fits}
\end{table}

For the coupling functional $U[N(r),n(r),V(r)]$ in (\ref{eq:JDFT1}),
we proceed by dividing it into two parts: long-range dielectric
screening $\Delta U_{sc}[n(r),V(r)]$ capturing the tendency of the
molecules in the liquid to be found in orientations and polarization
states that tend to screen
long-range electric fields, and a short-range electron-overlap
interaction $\Delta U_{el}[N(r),n(r)]$.  The long-range screening
depends only on the electrostatics of the solute and so depends only
on its electron density $n(r)$ and the nuclear electrostatic potential
$V(r)$.  The electron-overlap contribution depends upon contact
between the solvent molecules and the solute electron density and so
to some approximation depends only upon these two densities.

\begin{figure}
\includegraphics[width=3.25in]{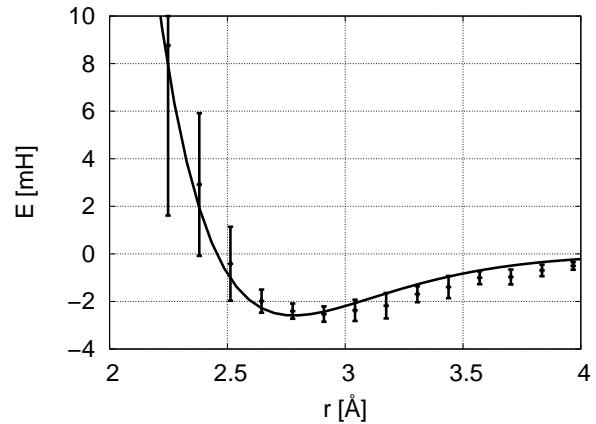}
\caption{Comparison of energy of interaction of water molecule with an
atom of neon as a function of neon-oxygen distance: results from
orientation independent pseudopotential (solid curve), orientally
averaged {\em ab initio} data (centers of error bars), typical
($1\sigma$) range of values as a function of orientation (range of
error bars).}
\label{fig:Vps}
\end{figure}

The lowest-order form for the electronic coupling between the liquid
and the solute which is compatible with translational and rotational
symmetry is then
$$
\Delta U_{el}[N(r),n(r)] = \int d^3R \int d^3r \,\, n(r) V_{ps}(r-R) N(R).
$$ Such a lowest-order coupling is a reasonable starting point as the
overlap is small.  With this form, the convolution kernel $V_{ps}$
plays the role of the average potential which an electron at point $r$
feels from a water molecule at point $R$, similar to the ``molecular''
pseudopotential of the type introduced by Cho and coworkers in the
different context described above\cite{Kim96}.  The main difference
between this potential and that of Ref.~\cite{Kim96} is that in the
present, preliminary formulation, the pseudopotential contains no
information about the orientation of the molecules and so represents
some sort of orientational average.  To optimize numerical convergence, we choose to fit $V_{ps}(r)$ to the sum of two origin-centered Gaussians of adjustable width and amplitude for a
total of four adjustable parameters,
$$
V_{ps}(r) = A_1 e^{-\frac{r^2}{2 \sigma_1^2}} + A_2 e^{-\frac{r^2}{2 \sigma_2^2}}
$$
 We then adjusted these
parameters to reproduce the orientationally averaged interaction of a
water molecule with an atom of neon as a function of distance.
Figure~\ref{fig:Vps} summarizes the results.
The parameters that we found are
$$
A_1 = 0.0765;
\sigma_1 = 2.045;
A_2 = -0.065;
\sigma_2 = 2.165.
$$ With this simple form, we were able to reproduce the average
interaction to within 1~millihartree ($\sim 0.63$~kcal/mole) for all
distances beyond 2\AA.  (Smaller distances are not very relevant as
the interaction becomes very repulsive.  For comparison, at room
temperature, $k_B T = 0.93$~mH.)  Here, there is no fitting to
empirical data, and the pseudopotential is truly {\em ab initio} in the same sense as an electron pseudopotential.
As such, Vps does not afford an opportunity to fold solvation data
into the construction of our functional.

Next, the screening term depends upon the long-range electrostatic
potential of the solute.  We take it to be
the change in electrostatic energy of this potential in the presence
of a linear, non-local
dielectric function of the form
$$\epsilon(r,r')\equiv
\delta(r-r')+\frac{4\pi\chi_b}{N_b^2} N(r) f(r-r') N(r'),
$$ where $\delta(r)$ is the Dirac-delta function, $f(r)$ is a
short-ranged function integrating to one, and $N_b$ and $\chi_b$ are
the density and static dielectric polarizabilities, respectively, of
the bulk liquid.  This choice ensures a smooth transition from the
dielectric constant of the bulk to that of vacuum over the
length-scale of $f(r)$.  The choice to include both $N(r)$ and $N(r')$
is motivated both by the need for $\epsilon(r,r')$ to be symmetric and
by the notion that the response at point $r$ to a field at point $r'$
depends on the density of molecules at both locations.  The connecting
function $f(r)$ was chosen to be a Gaussian of width
$\sigma$=2.25~bohr=$\sim$1.190~\AA, somewhat larger than the O-H bond
distance in water.  The motivation for using this value of $\sigma$ is
that, at lower values, dielectric screening at short length-scales is
so effective that the system may lower its energy by bringing fluid
density $N(r)$ into the atomic cores and the system thus becomes
numerically unstable.  We stress, however, that once stability was
achieved, the final results were not sensitive to the choice of this
parameter (typically 10\% variation in the solvation energy over the range
$\sigma=1.5$~bohr to $\sigma=2.5$~bohr) and that this parameter was in
no way adjusted to reproduce experimental solvation energies.
In the future, a direct description of the electric polarization in
terms of the orientational state of the solvent would capture
dielectric effects directly and remove the need for constructing such a
simple, {\em ad hoc} model.

\begin{figure}
\includegraphics[width=3.25in]{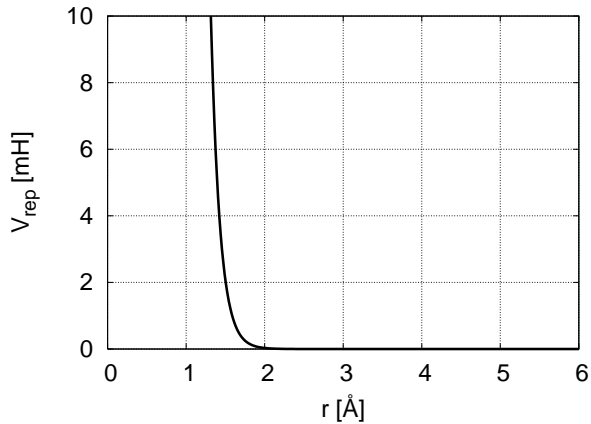}
\caption{Short-ranged repulsive potential added to prevent collapse of
liquid density $N(r)$ into the strong electric fields within the
atomic cores.  Once prevention of this artificial collapse is
achieved, the final results are insensitive to the choice of this
potential.}
\label{fig:Vrep}
\end{figure}

\begin{figure}
\includegraphics[width=3.25in]{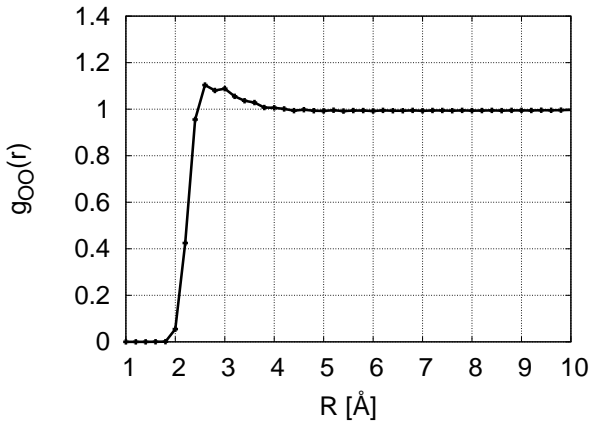}
\caption{Spherical average of $N(r)$ about the oxygen nucleus of an
explicit water molecule in solution scaled to the bulk liquid density,
corresponding to the correlation function $g_{OO}(r)$ measured in
experiments on liquid water.}
\label{fig:gOO}
\end{figure}

Finally, in conjunction with a nonlocal dielectric response, a term
must be added to the energy functional to help prevent the
aforementioned numerical issues associated with penetration of the
solvent density into the cores of the atoms.  To avoid this, we added
a short-ranged repulsive potential of the form $\int V_{rep}(r)
N(r)\,d^3r$ inside the atomic cores to prevent the overlap of the
solvent with the nuclei, where $V_{rep}(r)$ is taken to be a rapidly
decaying exponential function (rounded within $\sim$0.25~\AA~of the
origin) of constant 8.44~\AA$^{-1}$, leading to an apparent hard wall
at thermal energy scales near 1.5~\AA (Figure~\ref{fig:Vrep}).  Again,
once the system is numerically stable, the results are not sensitive
to the form of the repulsion (typically 1\% changes in solvation
energy) so long as the repulsion effectively cuts off before the
natural point of closest approach of the solvent at $\sim$2~\AA.  (See
Figure~\ref{fig:gOO} and the corresponding discussion below.)

{\em Results and Conclusions ---} 
The meaningful physical quantities predicted by joint-density functional
theory include not only the free energy $A$ but also the liquid
density $N(r)$.  Under certain circumstances, the later is accessible
directly in experiments.  For instance, when studying the solvation of
a water molecule with liquid water, the density $N(r)$
in our formulation gives the density of oxygen nuclei given a fixed
location for one water molecule.  As observed by Percus~\cite{Percus62},
the spherical average of $N(r)$ thus
corresponds to the oxygen-oxygen pair distribution function $g_{OO}(r)$
measured in experiments.  Figure~\ref{fig:gOO} shows our results for
this quantity.  In
good agreement with both x-ray and neutron scattering
experiments\cite{hura00}, we find $N(r)$ to be essentially zero until
a radius of $\sim$2.0~\AA, at which point the density rises rapidly,
overshooting the bulk density before finally approaching it.  The
experiments do show a much more pronounced peak and much more
structure in the form of oscillations which occur beyond 2.5~\AA.  We believe
that these discrepancies are due to the over simplified model we are using
for $A_{lq}[N(r)]$ in (\ref{eq:Alq}) and that a better liquid
density-functional would improve this.

Generally, the coordination
number for a liquid is defined as the integral of $g_{OO}$ from $r=0$
to the location of the first minimum after the coordination peak.  In
our case there are no such oscillations.  However, if we define the
coordination number as the integral of $g_{OO}$ up to 3.6~\AA, the
last point before the bulk value of unity is obtained, we find a
coordination of 5.3, relatively close to the value of approximately
four measured in liquid water and far from the close packed value near
twelve typically found in simple fluids.  Thus, our model, as simple
as it is, captures enough of the physics of water to reflect the
hydrogen bonding network which leads to tetrahedral coordination.
The model also appears to reflect the correct
energetics and to give a correct {\em ab initio} prediction of the
boundary position of the cavity outside of which the fluid is
excluded.  To our knowledge, this is the first accurate determination
of such a boundary directly from first principles.  As solvation
energies are known to be quite sensitive to the construction of the
boundary, this success gives a strong advantage to the current
approach.

\begin{figure}
\includegraphics[width=3.25in]{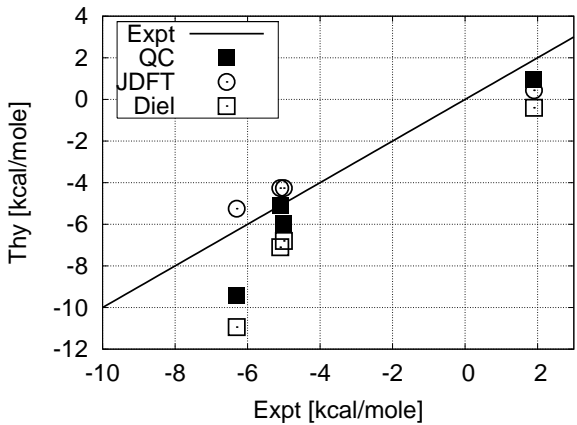}
\caption{Computed (vertical axis) versus experimental (horizontal
  axis) solvation energies for water, ethanol, methanol and methane
  (from left to right): perfect agreement (diagonal line), published
  quantum chemistry values (\cite{Marten} for all but water,
  \cite{Truong} for water) with dielectric contribution
  only (open squares) and including cavitation terms (closed squares), preliminary results from joint density functional theory (open circles).}
\label{fig:solvdat}
\end{figure}

Finally, Figure~\ref{fig:solvdat} summarizes the comparison of the
free energies predicted by the implementation described above with
both experiment and the predictions of state-of-the-art continuum
solvation models.  Our joint density-functional theory results
are clearly superior to dielectric-cavity-only calculations and are
arguably better than state-of-the-art continuum methods that include
cavity corrections.  It is satisfying to see that,
without any fitting or {\em ad hoc} adjustments, the hydrophobicity of
methane is predicted correctly.  It is also quite encouraging to see that the value we predict for the
most basic test, water in water, falls much closer than do the quantum
chemical methods to the correct value for inserting a water molecule
at a fixed location into liquid water.  While improvements and further
studies are needed, the quality of these results without adjustment of
parameters is very encouraging.

Although the results of the above preliminary implementation have the advantage of involving no direct fitting
to solvation energies and are of comparable quality to the
state-of-the-art, there are a number of weaknesses in the preliminary implementation, namely (1) the lack of
orientational dependence in the pseudopotential, (2) an artificial
separation of the dielectric response from the internal orientational
correlations of the liquid, (3) an over simplified, linear description
of the dielectric response with an {\em ad hoc} nonlocal length scale,
(4) the resulting need to artificially prevent the solvent from
overlapping the nuclei.  All of these can be addressed by a formulation including some information
describing orientational ordering in the fluid, which would allow for
the use of orientation-dependent pseudopotentials and could be used to
more naturally describe the dielectric response with its non-local
and non-linear effects.  Development of such an approach is the
subject of future work.

\end{document}